\documentclass[journal]{IEEEtran}

\ifCLASSINFOpdf
\else
   \usepackage[dvips]{graphicx}
\fi
\usepackage{url}
\usepackage{amsmath}
\usepackage{amssymb}
\usepackage{caption}
\usepackage{subcaption}
\usepackage{graphicx}
\usepackage{booktabs}
\usepackage{float}
\usepackage{multirow}
\usepackage{subcaption}
\usepackage{array}
\newcolumntype{P}[1]{>{\centering\arraybackslash}p{#1}}
\newcolumntype{M}[1]{>{\centering\arraybackslash}m{#1}}
\renewcommand{\thetable}{\Roman{table}}
\usepackage{hyperref}

\usepackage{tikz}

\newcommand\submittedtext{%
  \footnotesize This work has been submitted to the IEEE for possible publication. Copyright may be transferred without notice, after which this version may no longer be accessible.}

\newcommand\submittednotice{%
\begin{tikzpicture}[remember picture,overlay]
\node[anchor=south,yshift=10pt] at (current page.south) {\fbox{\parbox{\dimexpr0.65\textwidth-\fboxsep-\fboxrule\relax}{\submittedtext}}};
\end{tikzpicture}%
}

\begin{document}
\title{End-to-End Target Speaker Speech Recognition Using Context-Aware Attention Mechanisms for Challenging Enrollment Scenario}
\author{Mohsen Ghane, Mohammad Sadegh Safari
\thanks{ This work was conducted at the Part Artificial Intelligence Research Center.}
\thanks{M. Ghane is a master's student at Aalto University, Finland (e-mail: mohsen.ghane@aalto.fi). He was affiliated with the Part Artificial Intelligence Research Center during the course of this research}
\thanks{M. S. Safari is with  Part Artificial Intelligence Research Center (mohammadsadeq.safari@partdp.ai).}}

\markboth{}{}
\maketitle

\begin{abstract}
This paper presents a novel streaming end-to-end target-speaker speech recognition that addresses two critical limitations in systems: the handling of noisy enrollment utterances and specific enrollment phrase requirements. This paper proposes a robust Target-Speaker Recurrent Neural Network Transducer (TS-RNNT) with dual attention mechanisms for contextual biasing and overlapping enrollment processing. The model incorporates a text decoder and attention mechanism specifically designed to extract relevant speaker characteristics from noisy, overlapping enrollment audio. Experimental results on a synthesized dataset demonstrate the model's resilience, maintaining a Word Error Rate (WER) of 16.44\% even with overlapping enrollment at $-\text{5dB}$ Signal-to-Interference Ratio (SIR), compared to conventional approaches that degrade to WERs above 75\% under similar conditions. This significant performance improvement, coupled with the model's semi-text-dependent enrollment capabilities, represents a substantial advancement toward more practical and versatile voice-controlled devices.
\end{abstract}
\submittednotice

\IEEEpeerreviewmaketitle

\section{Introduction}

\IEEEPARstart{A}{utomatic} Speech Recognition (ASR) systems have made significant advancements, especially in voice-controlled devices like smart speakers. However, they still encounter challenges in real-world conditions, particularly with overlapping speech and noisy environments. Accurately recognizing speech from a target speaker in the presence of interference, such as the noise from a TV or  background conversations, is essential for the effectiveness of these devices.

Blind Speech Separation (BSS) is a widely used approach to extract individual speech signals from mixed audio without prior knowledge of the speakers \cite{Jadhav2008}. Earlier work proposed methods, such as matrix factorization and signal decomposition using probabilistic generative models and tree-structured graphical models. These approaches incorporate prior knowledge about latent variables or sources, such as spatio-temporal decorrelation, statistical independence, sparseness, and smoothness \cite{Choi2004, Hild2008, Cichocki2006}. Notable methods include Deep Clustering, which uses spectrogram embeddings for class-independent signal separation \cite{hershey2015}, and Conv-TasNet, a time-domain audio separation method that employs a temporal convolutional network (TCN) with stacked 1-D dilated convolutional blocks for long-term dependency modeling in a compact size \cite{luo2019conv}. Additionally, the Resource-Efficient Separation Transformer is a Transformer-based model designed to reduce the computational cost of speech separation tasks \cite{Libera2024}. Despite promising results, three key limitations could make them less suitable for voice-controlled devices. Firstly,  severe performance degradation in streaming scenarios, which is a crucial need for real-time devices \cite{Veluri2022}. Secondly, dependence on prior knowledge of the number of speakers in the mixture, which is often unavailable in practice \cite{Choi2004}. Finally, the production of multiple outputs requires additional target speaker information for identification.

Targeted speech separation is a technique that isolates a target speaker's voice from mixed audio signals using auxiliary information like enrollment utterances \cite{molkov2023}. Speaker Beam \cite{Zmolikova2019} and VoiceFilter \cite{Wang2019}, for instance, employ a speaker embedding network that relies on a pre-trained model to generate a speaker embedding from an enrollment utterance to condition a neural network, typically a recurrent neural network (RNN), for speech extraction. While these networks act as a preprocessing step on the input of ASR models, there is a notable shift towards end-to-end models in target-speaker speech recognition, aiming to streamline the entire process from raw audio input to separated speech output. End-to-end models offer several advantages, including reduced complexity and improved performance by optimizing the entire pipeline jointly \cite{Wakayama2024, He2021, Ochiai2020}. While the proposed end-to-end models demonstrate state-of-the-art performance, they are limited by one of the two major drawbacks. First, enrollment audio must be relatively clean, with the target speaker as the dominant source of sound  \cite{Moriya22, Ji2020, Shi2020, Raj2023}. To overcome this, one possible solution is to use text-dependent models that should be trained on a dataset specifically created for a particular enrollment phrase (for example, "Hey, computer") \cite{Wang2019}, limiting adaptability if a different wake word is needed. These limitations constrain the practical deployment of end-to-end target-speaker speech recognition models in real-world scenarios, where users might not have control over the quality of the enrollment audio or may prefer to use different enrollment phrases.

In this paper, we aim to address these limitations by building upon the TS-RNNT framework. In Section \ref{ts-rnnt}, we introduce the TS-RNNT and related foundational concepts, providing an overview of its architecture and capabilities. In Section \ref{propsed}, we offer potential contributions to this model by: 
\begin{itemize}
\item  Enhancing the capability of handling short, noisy, and even overlapping enrollment utterances. This advancement significantly enhances the system's performance in real-world scenarios, where clean enrollment data is often unavailable or impractical to obtain.
\item Achieving semi-text-dependence in the enrollment phase, offering unprecedented flexibility by allowing for adaptation to various phrases without the need for specialized training data.
\end{itemize}
These contributions address critical limitations in existing models, making it easier to create practical and flexible voice-controlled devices. Our model's ability to operate effectively in challenging acoustic environments while maintaining flexibility in enrollment requirements. 



\section{Streaming TS-RNNT}\label{ts-rnnt}

\begin{figure}[!h]
    \centering
    \begin{subfigure}[b]{0.35\columnwidth}
        \centering
        \includegraphics[width=\textwidth]{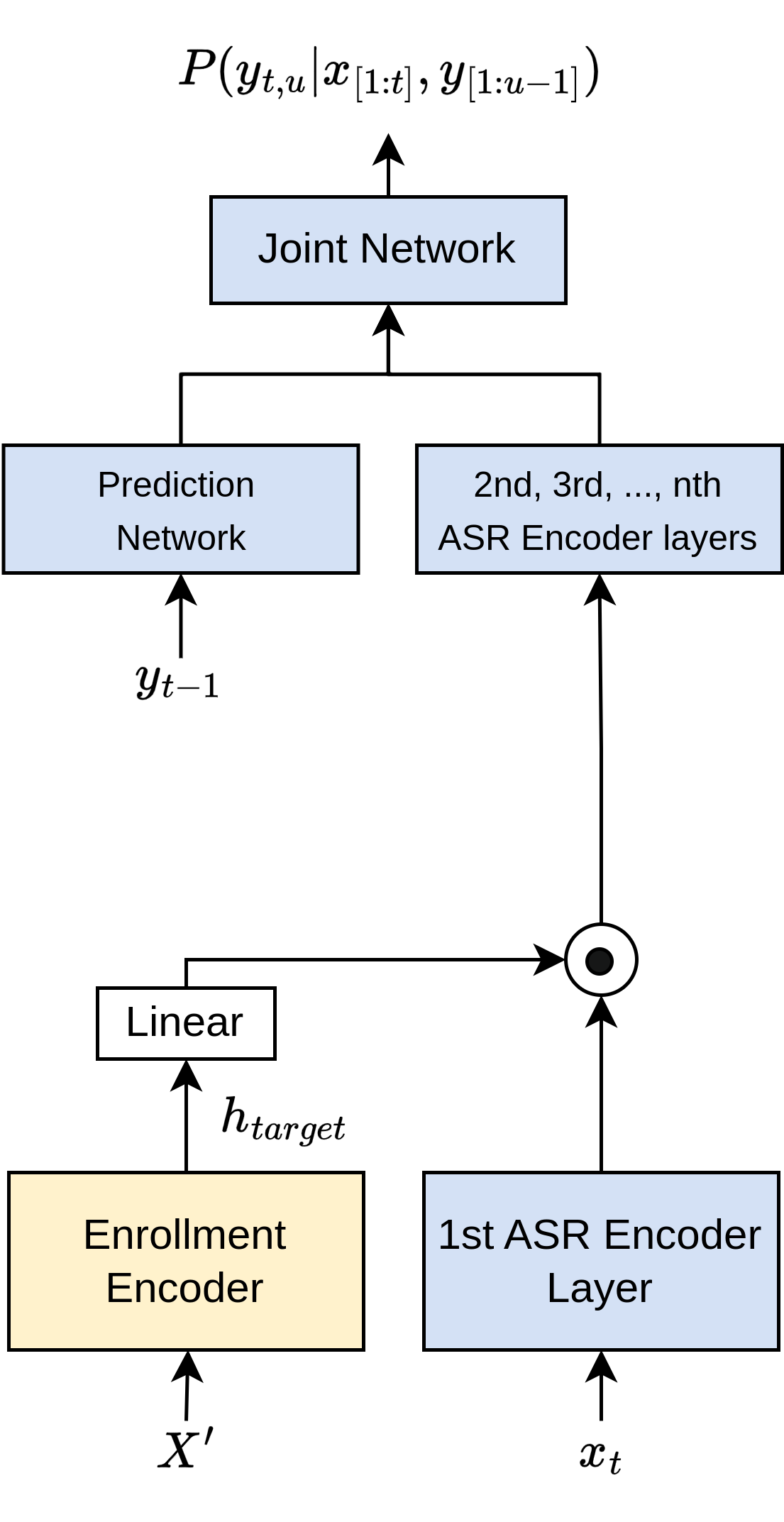}
        \caption{\fontsize{8}{8}\selectfont TS-RNNT}
        \label{fig:subfig1}
    \end{subfigure}
    \hfill
    \begin{subfigure}[b]{0.55\columnwidth}
        \centering
        \includegraphics[width=\textwidth]{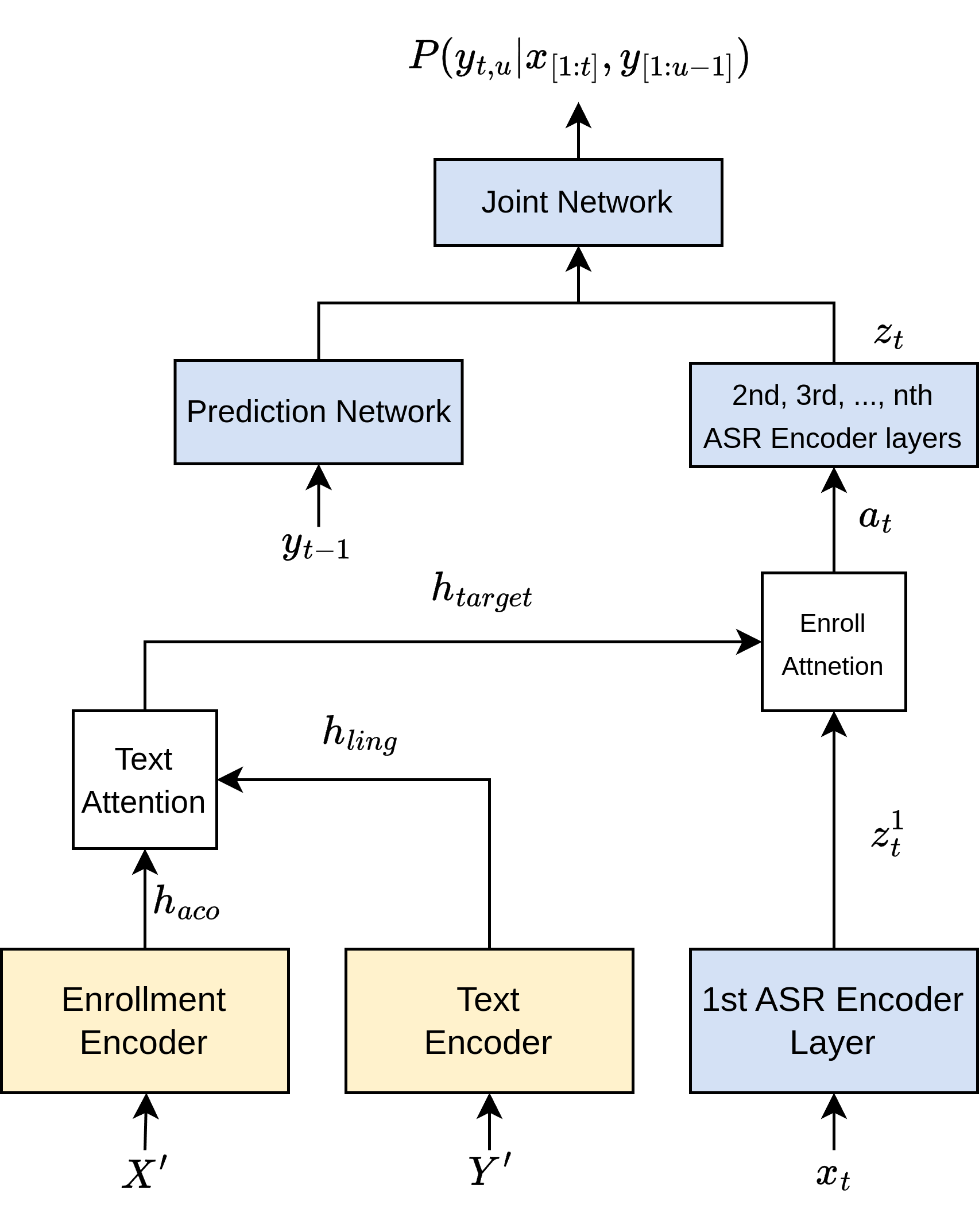}
        \caption{\fontsize{8}{8}\selectfont RobustTS-RNNT}
        \label{fig:subfig2}
    \end{subfigure}
    \label{fig:mainfig}
    \caption{Architectures of the Target-Speaker RNN Transducer (TS-RNNT). In (a), the TS-RNNT architecture includes standard RNNT components (blue), modules that capture target speaker characteristics (yellow), and speaker-specific bias (white). In (b), the RobustTS-RNNT architecture extends TS-RNNT by incorporating Contextual Biasing and a Text Attention Mechanism, where additional yellow modules capture target speaker features with text information. Model weights in RobustTS-RNNT are initialized from the offline Zipformer (yellow) and streaming Zipformer (blue).}
\end{figure}

Streaming Target-Speaker Recurrent Neural Network Transducer (TS-RNNT), introduced by Moriya et al.~\cite{Moriya22}, is an end-to-end approach for target-speaker ASR that supports streaming recognition. TS-RNNT architecture integrates target-speaker speech recognition functionality within a standard RNNT framework (see \autoref{fig:subfig1}). It consists of four main components: an enrollment encoder, an ASR encoder, a prediction network, and a joint network.

The enrollment encoder, denoted as $f_\text{Enr-Enc}$, processes the enrollment audio of the target speaker $X'$ to produce a speaker embedding $h_\text{target}$. This embedding captures the unique characteristics of the target speaker's voice. The ASR encoder, $f_\text{ASR-Enc}$, is the core component that enables target-speaker speech recognition. It takes as input both the mixed speech signal $X$ and $h_\text{target}$. First, the embedding goes through a linear layer and is averaged over time. Then, it integrates into the ASR process by combining with an intermediate layer of the encoder (typically the first layer, as this has been shown to be most effective) using a simple element-wise multiplication, also known as a Hadamard product. This fusion allows the encoder to focus on the target speaker's voice within the mixed input. The prediction network, $f_\text{Pred}$, and the joint network, $f_\text{Joint}$, function similarly to those in a standard RNNT. The prediction network generates token representations based on previous predictions, while the joint network combines the outputs of the ASR encoder and prediction network to produce token posterior probabilities.

To enable streaming capability, the ASR encoder employs a left-to-right encoding scheme to enable real-time, low-latency processing of input speech, making the system suitable for streaming applications. The speaker encoder remains unchanged, as it computes the speaker embedding from short enrollment audio, such as a wake word, with minimal latency. During inference, the system first calculates the speaker embedding from the enrollment audio. Each frame of mixed speech is then encoded, fused with the speaker embedding, and processed to generate token probabilities, which are finally decoded using the RNNT framework to produce the transcription.



\section{Proposed method} \label{propsed}
The Streaming Target-Speaker Recurrent Neural Network Transducer (TS-RNNT) has shown strong performance but faces two key limitations. First, it uses a single static bias vector, applied uniformly across all time frames, to represent the target speaker. This approach fails to capture temporal variations in the input and performs poorly with short enrollment audio, where limited information results in a less effective representation of the target speaker. Second, TS-RNNT struggles in overlapping enrollment scenarios where competing speech is present. Without mechanisms to isolate the target speaker's features from interference, the model cannot effectively differentiate the target speaker’s characteristics from those of other speakers, resulting in poor recognition accuracy in such conditions.

To address these issues, we propose a Robust Target-Speaker Recurrent Neural Network Transducer. Our method introduces dynamic contextual biasing to improve the representation of the target speaker in short enrollment audio and incorporates text-guided attention mechanisms to handle overlapping enrollment scenarios. These enhancements enable the model to generate adaptive and refined speaker embeddings, improving robustness and accuracy in real-world conditions.

\subsection{Contextual Biasing}
\label{attention}
Overcoming the limitations of static bias vectors in TS-RNNT requires a more adaptive approach to effectively capture the nuances of the target speaker across varying temporal contexts. To address this, an attention mechanism~\cite{vaswani2023} is employed to generate context-dependent bias vectors that adapt dynamically over each time frame, ensuring that the unique characteristics of the target speaker~\cite{Xiao2019} are effectively captured without losing critical information from the data. The attention unit, $f_\text{Enr-Att}$, is added to the model to implement this functionality. In this configuration, $f_\text{Enr-Enc}$ output is utilized as both the key and value in $f_\text{Enr-Att}$, while the intermediate ASR layer serves as the query. Consequently, the computation of $f_\text{ASR-Enc}$ output can be expressed as follows:

\begin{align}
    & h_\text{target} = f_\text{Enr-Enc}(X') \label{h_target}\\
    & z_t^1 = f_\text{ASR-Enc}^1(x_t) \nonumber \\
    & a_t = f_\text{Enr-Att}(z_t^1,h_\text{target},h_\text{target})\nonumber \\
    & \phantom{a_t} =  \text{softmax}(\frac{z_t^1 h_\text{target}^T} {\sqrt{d_h}})h_\text{target}  \nonumber\\
    & z_t = f^{n-1}_\text{ASR-Enc}(a_t)  \nonumber
\end{align}

where $f_\text{ASR-Enc}^{1}$ represents the first layer of the ASR encoder, while $z_t^1$ denotes the output generated by this first layer at time frame $t$. The variable $a_t$ is calculated as the attention output for time frame $t$. Finally, $f_\text{ASR-Enc}^{n-1}$ refers to the remaining layers of the ASR encoder, and $z_t$ is the final output of the ASR encoder at time frame $t$.  Additionally, $h_\text{target} \in \mathbb{R}^{T \times d_h}$, where  $T$ is the number of time frames and $d_h$ is the number of acoustic features.


\subsection{Text-Guided Attention}
To further improve the robustness of speaker embeddings, especially in scenarios involving overlapping enrollment audio, a text-guided attention mechanism is introduced into the model, as illustrated in \autoref{fig:subfig2}. This mechanism consists of a text decoder $f_\text{Txt-Dec}$  and an additional attention unit $f_\text{Txt-Att}$. These components enable the extraction of speaker embeddings that are specifically focused on the target speaker, by leveraging linguistic information from predefined wake words or enrollment text.

Since smart devices are activated with pre-defined wake words the model knows what the target speaker is saying. Using that information, during the speaker embedding phase, the model decodes the corresponding wake word's text $Y'$. Subsequently, the output of $f_\text{Enr-Enc}$ is processed through $f_\text{Txt-Att}$ as the query, where $f_\text{Txt-Dec}$ output acts as both the key and value. In the proposed model $h_{target}$ in equation \ref{h_target} can be formulated as follows:

\begin{alignat}{2}
    & h_\text{aco}       &{}={}& f_\text{Enr-Enc}(X')\nonumber \\ 
    & h_\text{ling}       &{}={}& f_\text{Txt-Dec}(Y')\nonumber \\ 
    & h_\text{target}  &{}={}& f_\text{Txt-Att}(h_\text{aco},h_\text{li},h_\text{ling}) \nonumber \\
    &                  &{}={}& \text{softmax}\left(\frac{h_\text{aco} h_\text{ling}^T}{\sqrt{d_l}}\right) h_\text{ling} \label{text-Attention}
\end{alignat}

Where $h_\text{aco} \in \mathbb{R}^{T \times d_{a}}$ is the acoustic embedding of the enrollment, with $T$ representing the number of time frames and $d_{a}$ denoting the number of acoustic features. Additionally, $h_\text{ling} \in \mathbb{R}^{N \times d_{l}}$ is the linguistic embedding of the text, with $N$ representing the number of tokens and $d_{l}$ denoting the number of linguistic features.

The equation \ref{text-Attention} prioritizes acoustic features representing $Y'$ tokens and ignoring irrelevant features. As a result, the bias vector $h_\text{target}$ focuses only on the speaker who said $Y'$, without considering any environmental or speech noise. This architecture is semi-text-dependent, as $f_\text{Txt-Dec}$ and $f_\text{Txt-Att}$ can process any given wake word $Y'$, rather than being constrained to a single phrase as in conventional text-dependent systems. While the model requires text guidance to generate $h_\text{ling}$, subsequently influencing $h_\text{target}$, it can process any wake word during both training and inference.



\begin{table*}[h!]
    \centering
    \resizebox{\textwidth}{!}{
    \begin{tabular}{@{}M{2.8cm} M{2cm} ccccccccccc@{}} 
        \toprule
        \multirow{2}{*}{Model} & \multirow[M]{2}{=}{Overlapping Enrollment} & \multicolumn{11}{c}{SIR Level} \\ 
        \cmidrule(lr){3-13} 
        &  & 5 & 4 & 3 & 2 & 1 & 0 & -1 & -2 & -3 & -4 & -5 \\ 
        \midrule
        \multirow{2}{*}{TS-RNNT} & 
        \texttimes & 13.05 & 13.4 & 13.8 & 14.17 & 14.68 & 15.19 & 
        15.64 & 16.13 & 16.74 & 17.63 & 18.55 \\ 
        & 
        \checkmark & 26.94 & 30.62 & 35.58 & 40.37 & 46.19 & 51.88 & 
        57.85 & 62.86 & 68.05 & 72.14 & 75.34 \\ \midrule
        
        \multirow{2}{*}{AttentiveTS-RNNT} & 
        \texttimes & 11.5 & 11.74 & 11.93 & 12.27 & 12.58 & 12.89 & 
        13.21 & 13.67 & 14.03 & 14.59 & 15.27 \\ 
        & 
        \checkmark & 23.67 & 27.57 & 33.08 & 38.22 & 44.36 & 51.3 & 
        58.18 & 64.16 & 69.27 & 74.29 & 78.04 \\ \midrule
        
        \multirow{2}{*}{RobustTS-RNNT} & 
        \texttimes & 12.1 & 12.4 & 12.6 & 12.85 & 13.18 & 13.46 & 
        13.82 & 14.22 & 14.69 & 15.2 & 15.78 \\ 
        & 
        \checkmark & 12.7 & 13.01 & 13.3 & 13.6 & 13.94 & 14.21 & 
        14.58 & 15.04 & 15.37 & 15.95 & 16.44 \\ 
        \bottomrule
    \end{tabular}
    }

    \caption{The table presents the evaluation results of various models under controlled noise conditions. Each model was tested with audio samples containing overlapping speech at different Signal-to-Interference Ratios (SIR), with the enrollment either including or excluding overlapping speech. Additionally, background noise at SNR levels ranging from 0 dB to 20 dB was introduced into each test audio. For all models, ASR encoder is a streaming large Zipformer, and the enrollment encoder is an offline small Zipformer.}
    \label{tab:model_performance}
\end{table*}

\section{Expriments}
\subsection{Dataset}
Considering that there is no existing dataset that explicitly addresses target-speaker speech recognition with overlapping enrollment, a new dataset was synthesized by combining two existing datasets. The first dataset, \emph{DeepMine} \cite{zeinali18b}, includes over 1,300 speakers and approximately 500 hours of Persian (Farsi) automatic speech recognition (ASR) training data. The second dataset,\emph{WHAM!} \cite{wichern2019}, contains approximately 80 hours of noise backgrounds recorded with a binaural microphone across 44 different locations, including coffee shops, restaurants, and bars.

Each sample is generated by selecting two random samples from different speakers and a noise sample from \emph{WHAM!}. To simulate real-world scenarios, the following procedure is done for each sample generation:

\begin{enumerate}
  \item Two utterances and the noise are reverberated by different sources located in one room, following the methodology described by \cite{ko2017}.
  
  \item Using the \emph{Montreal Forced Aligner} \cite{McAuliffe2017}, each utterance is divided into two parts: enrollment and command. The enrollment part is the initial segment of the speech, lasting up to 1.5 seconds, while the command part includes the rest of the audio. 

  \item The enrollment segments from both speakers are combined with a random Signal-to-Interference Ratio (SIR) between $[-5,5]\space \text{dB}$, and the same is done for the command segments with the same SIR as the enrollment mixture.

  \item The first 1.5 seconds of the noise are combined with the enrollment part, and the rest of the noise is combined with the command part, both at the same Signal-to-Noise Ratio (SNR) ranging from $[0,20]\space \text{dB}$.
\end{enumerate}

\subsection{Zipformer}
Zipformer is a novel audio encoder for automatic speech recognition, that offers significant advancements in performance, memory efficiency, and computational speed compared to other models, such as Conformer and Squeezeformer \cite{yao2024}. Its U-Net-like structure allows middle stacks to operate at lower frame rates, reducing computational demands without compromising accuracy. Additionally, Zipformer's restructured blocks, which include more modules and reuse attention weights, enhance performance while maintaining lower memory usage and faster inference times. The ScaledAdam optimizer, unique to Zipformer, ensures faster convergence and superior performance compared to the traditional Adam optimizer.

This balanced trade-off between performance and efficiency, with reduced computation time and memory requirements, made Zipformer particularly suited for voice-controlled devices with constrained hardware, where real-time performance is critical. In this work, Zipformer is used for both Enrollment and ASR encoding tasks.

\subsection{Training}
Our model's training followed a three-stage approach. Initially, we initialized the model using weights from two pre-trained regular TS-RNNTs: an offline small Zipformer and a streaming medium Zipformer, both trained on command portions of the dataset. The ASR encoder, prediction network, and joint network were weighted by the streaming Zipformer, while the enrollment encoder and text encoder (which shares the same architecture as the prediction network in the offline Zipformer) were weighted by the offline Zipformer (see \autoref{fig:subfig2}). Both attention mechanisms used an embedding dimension of 256 and 8 attention heads. The second stage involved training on the dataset without overlapping enrollments using the ScaledAdam optimizer with a learning rate of 0.01, batches containing a maximum of 900 seconds of audio data, and running for 150 epochs. Finally, we fine-tuned the model on the overlapping enrollment dataset, reducing the learning rate to 0.005, increasing the maximum batch duration to 1200 seconds, and training for 200 epochs. Across all stages, standard RNN-T loss was employed for end-to-end optimization. This progressive training strategy allowed the model to establish a strong baseline before adapting to the complex task of handling overlapping enrollments. 

\subsection{Results}
We evaluated several variants of our model architecture (see Table \ref{tab:model_performance}). TS-RNNT serves as our baseline, using simple mean pooling to combine speaker embedding with the ASR encoder output. AttentiveTS-RNNT extends this by incorporating an attention mechanism for context-dependent speaker biasing. Finally, RobustTS-RNNT represents our complete architecture, adding a text decoder and attention mechanism specifically designed for handling overlapping enrollment scenarios.

Attentive TS-RNNT outperformed the baseline TS-RNNT in scenarios without overlapping enrollment, achieving a lower WER (11.5\% compared to 13.05\% at 5dB SIR). This improvement can be attributed to the introduction of an attention mechanism that enables dynamic contextual biasing, effectively capturing nuanced speaker characteristics and addressing the limitations of static bias vectors. However, the significant advantages of our proposed architecture become apparent when dealing with overlapping enrollment. While both TS-RNNT and AttentiveTS-RNNT showed acceptable performance without overlapping enrollment, their performance degraded dramatically with overlapping enrollment, reaching WERs above 75\% at -5dB SIR. In contrast, RobustTS-RNNT maintained remarkable stability, achieving a WER of 16.44\% with overlapping enrollment at -5dB SIR, only slightly higher than its non-overlapping performance of 15.78\%. This demonstrates the effectiveness of our enrollment decoder and attention mechanism in isolating target speaker characteristics under highly challenging conditions.

\section{Conclusion}

This paper has introduced a robust approach to target-speaker speech recognition that significantly advances the state of the art in handling challenging real-world scenarios. Our proposed RobustTS-RNNT architecture, with its dual attention mechanisms and text decoder, demonstrates remarkable resilience to overlapping enrollment conditions while maintaining high transcription accuracy. The experimental results show that our model maintains consistent performance across various SIR levels, with only minimal degradation in highly challenging conditions (16.44\% WER at -5dB SIR with overlapping enrollment), whereas conventional approaches fail dramatically. This work attempts to address some limitations in existing systems by handling noisy, overlapping enrollment utterances and enabling semi-text-dependent enrollment, which may contribute to the development of more practical voice-controlled devices for real-world environments.

\bibliographystyle{IEEEtran}
\bibliography{refs2024}

\end{document}